\documentclass[%
 reprint,
 amsmath,amssymb,
 aps,
]{revtex4-2}

\usepackage{graphicx}% Include figure files
\usepackage{dcolumn}% Align table columns on decimal point
\usepackage{bm}% bold math
\usepackage{subcaption}
\usepackage{float}
\begin{document}

\preprint{APS/123-QED}

\title{Complex Networks characterization of Indian Water Dam Systems and its topographic response}% Force line breaks with \\
%\thanks{A footnote to the article title}%

\author{Dhruv Patel}
 %\altaffiliation{Department of Physics, Indian Institute of Science Education and Research, Pune, India}%Lines break automatically or can be forced with \\
% \author{Second Author}%
%  \email{Second.Author@institution.edu}
\affiliation{Department of Physics, Indian Institute of Science Education and Research, Pune, India%
 \\
}%

%\collaboration{MUSO Collaboration}%\noaffiliation

% \author{Charlie Author}
%  \homepage{http://www.Second.institution.edu/~Charlie.Author}
% \affiliation{
%  Second institution and/or address\\
%  This line break forced% with \\
% }%
% \affiliation{
%  Third institution, the second for Charlie Author
% }%
% \author{Delta Author}
% \affiliation{%
%  Authors' institution and/or address\\
%  This line break forced with \textbackslash\textbackslash
% }%

% \collaboration{CLEO Collaboration}%\noaffiliation

%\date{\today}% It is always \today, today,
             %  but any date may be explicitly specified

\begin{abstract}
\noindent
In this paper a Complex Network approach is taken to understand the salient features of Indian Water Dam Networks. Detailed analysis of 15 river basin networks have been carried out. The data has been taken from "River Basin Atlas of India" compiled by the Indian Space Research Organisation (ISRO) and Central Water Commission (CWC), Ministry of Water Resources, Government of India. The paper also investigates the correlation between various structural properties of the networks like total number of nodes, Link Density, Clustering Coefficient amongst each other and also with the Irrigation Potential and topographical features like the Elevation gradient of the region measured in meters. A mathematical model has also been proposed to understand the relation between irrigation potential measured in thousand hectares unit with the number of nodes, i.e. dams and barrages, to get a more quantitative understanding of the system. The paper also tries to observe the response of the network properties to actual topographical features of the region. This lays down a basic foundational work in understanding these water dam networks through a complex network approach over which further work can be done to make the predictions more efficient. 
% \begin{description}
% \item[Usage]
% Secondary publications and information retrieval purposes.
% \item[Structure]
% You may use the \texttt{description} environment to structure your abstract;
% use the optional argument of the \verb+\item+ command to give the category of each item. 
% \end{description}
\end{abstract}

%\keywords{Suggested keywords}%Use showkeys class option if keyword
                              %display desired
\maketitle

%\tableofcontents

\section{\label{sec:level1}INTRODUCTION}
\noindent
Recently, the use of Complex Networks to understand systems with multiple components interconnected with each other have increased because of the simplicity, efficiency and the understanding of the system properties one gets using this approach. A lot of these properties are also dependent on the structures of these networks and the way they are connected\cite{Watts1998}. A lot of studies have been done and it is currently going on in the field of Complex Networks and understanding its structural properties\cite{doi:10.1126/science.286.5439.509, RevModPhys.74.47, doi:10.1137/S003614450342480, doi:10.1080/00018730110112519, BOCCALETTI2006175}.
\par
\noindent
Since the nodes of the networks are interconnected with each other, sometimes controlling a few nodes can change the output of the system a particular network represents. Often, the nodes which form hubs in the networks can be used to control the overall properties of the network.\cite{Sanhedrai2023,Sanhedrai2022}. Along with the advantages this provide, there are also serious consequences when one of these nodes fail to function and the failure propagates throughout the network, these failures can be intentional or non-intentional\cite{Smolyak2020,10.1093/comnet/cnaa013,PhysRevLett.86.3682}. Often, scale-free networks\cite{RevModPhys.74.47} show resilience to these failures, but they too are threatened in case of any targeted attacks\cite{Albert2000}.Thus, studying the network structures is really important to develop resilience towards such failures and maximizing the output we require from the system. In case of the Indian Water Dam Networks, because cascading failures or dis-functioning of a single node can affect lives of a lot of people. Even major failures can lead to adverse consequences on the agricultural, economical and health conditions of the people in the affected area. So understanding the stability of network becomes very essential, so that measures can be taken to prevent such incidents. 
\par
\noindent
The Complex network approach has been used to understand various such systems, such as electrical power grid networks \cite{Tamhane_2023,9521783,FORSBERG2023129072}, Water Distribution Systems \cite{10.1063/1.3540339,https://doi.org/10.1029/2020WR027929,SITZENFREI2021117359,w15081621} and many other physical, biological and social systems.\cite{article,10.1093/acprof:oso/9780199206650.001.0001,10.5120/ijca2015905952}. And the use and importance of complex networks have encouraged researchers to study different network properties of the system. Some of these networks are physical networks, which puts additional constraints on its structures and overall characteristics.\cite{PhysRevE.73.066107,doi:10.1061/(ASCE)0733-9496(2005)131:1(58)}. In case of Water Dam Networks taking into account the topographical, geographical and structural conditions is necessary to understand the stability of the network efficiently. It is also equally important while making strategic decisions relating to building of additional dams or barrages so that the overall output in terms of water management and irrigation is maximized. And the Water Dam Networks remain robust against any failure after these infrastructure projects have been implemented. There have been recent developments in studying such systems.\cite{CRUCITTI200492,Buhl2006,Masucci2009,10.1063/1.3540339}

\par
\noindent
In the paper, the first section starts with stating the information about the Indian Water Dam Networks and how they are formally organized by the Government of India and the Indian Space Research Organisation. The Appendix of the paper contains the network created for all the 15 river basins that have been studied in the paper. After this, the Brahmaputra river basin network has been used as an illustration for the network analysis that has been done on all the river basin networks. The Structural Properties of these Networks have been investigated in the next section. Here the Link Densities, total nodes, Clustering coefficient,etc have been enlisted for these networks and analyzed to derive useful information. Henceforth, more practical output of these Water Dam Networks, that is, the Irrigation potential they carry have been studied. The paper concludes with trying to understand the response of these networks to the topography of the river basin region. 
\vspace{0.5cm}

\section{STRUCTURE OF INDIAN WATER DAM NETWORKS}
\noindent
The Indian subcontinent consists of a large number of rivers which are significant both for the social, cultural, and, economic development of the country. Many major cities of India are located along these rivers. India inhabits 17.84\% of the total world population across 2.4\% of the world geographical area  and sustains it through sharing 4\% of the world’s fresh water resources\cite{India-WRIS.2012}. India is an agrarian country, with a major chunk of its population living in rural areas and it has the highest irrigated land in the world. These people are dependent on agricultural activities for their livelihood and hence, the river water resources are exploited for irrigation, generation of hydro-power and water supply. Due to rapid rise in the population after independence, has led to increasing food demand, stable and developing economy and improvement in standard of living; and all these factors have led to increased pressure on available natural water resources. Thus, proper management, planning and development of these water resources is pivotal for the growth of the country.
\par
\noindent
Optimization and management of these water resources requires a multidisciplinary perspective, with people from different areas of expertise like Science and Technology, Commerce, Strategic planning and Management participating to get the best possible results. Use of Complex Networks to analyze the properties of such systems have led to predictable and useful information which can be taken into consideration while designing projects for development for these systems and also for policy making. River basins are ideal units for planning and implementation of water resources projects. They provide ecologically sound and economically cost effective solutions for development and conservation. Basins have defined water boundaries within which there is an interrelationship between the surface and groundwater resources and provide a basis for planning overall development activities. The basin planning also presents comprehensive development possibilities of land and water resources to meet the anticipated regional and local needs\cite{India-WRIS.2012}. 
\par
\noindent
The river system in India is classified into four groups - Himalayan Rivers, Deccan Rivers, Coastal Rivers and River of Inland Drainage\cite{India-WRIS.2012}.  The  country is divided into 25 river basins. Here, I have worked with 15 major river basins of these 25. It includes Indus Basin, Ganga Basin, Brahmaputra Basin, Barak and other, Godavari Basin, Krishna Basin, Cauvery Basin, Subarnarekha Basin, Brahmani and Baitarani Basin, Mahanadi Basin, Pennar Basin, Mahi Basin, Sabarmati Basin, Narmada Basin, and, Tapi Basin. The information used in this paper for making the networks for these basins and doing other related analysis for it like the Irrigation Potential and geographical topology has been derived from the “ River Basin Atlas of India” . ‘The River Basin Atlas of India’ is an outcome of a collaborative effort between Central Water Commission (CWC), Government of India, and the Indian Space Research Organisation (ISRO) for the joint project ‘Generation of Database and implementation of Web Enabled Water Resources Information System \href{https://indiawris.gov.in/wris/#/atlas}{(India-WRIS)}‘. In the supplementary section at the end, you can see the networks of Water Dams and Barrages for these 15 different basins along the rivers.

\section{ILLUSTRATION}
\noindent
In this section, the network properties of Brahmaputra river basin have been analyzed in detail. The same analysis is done for the rest of the river basin networks as well and the network properties are summarized in the next section. The supplementary section of the paper contains the network structures for all the 15 river basin networks of India.

\begin{figure}[h]
  \includegraphics[width=\linewidth]{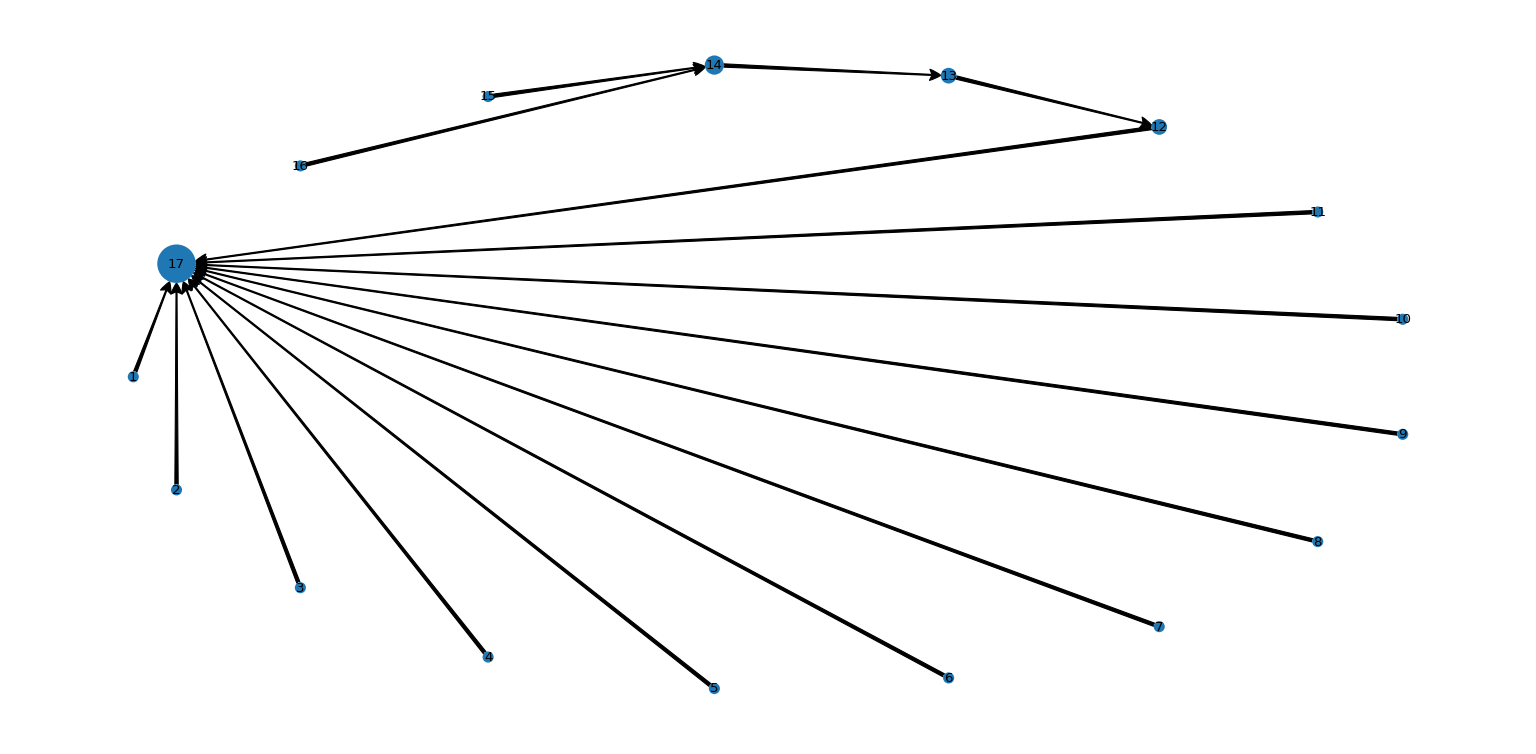}
  \caption{Brahmaputra River Basin Network.}
  \label{fig:1}
\end{figure}

\par
\noindent
The hub in the network, i.e. the node with the most number of incoming links, is the node that represents the Bay of Bengal. Other nodes represent a dam or a barrage. Here using this network for Brahmaputra river basin, the network properties like the Clustering Coefficient, Link Density, total number of nodes, Elevation Gradient and the Irrigation Potential. Investigating correlations between any of these properties across all the networks has been the goal of this paper and is carried out in the later sections.

\par
\noindent
The Irrigation Potential, measured in Thousand Hectares units, is calculated for the network by summing up the Irrigation Potential of the individual dams in the network or the irrigation projects corresponding to the dams. The total sum comes out to be 703.77 Th Ha, which we consider to be the Irrigation Potential for the Brahmaputra River Basin network The information regarding this is provided in the table \ref{Table:1} below.

\begin{table}[h]
\begin{tabular}{|l|l|}
\hline
Name of Project        & Irrigation Potential \\
\hline
Teesta Barrage         & 527.00               \\
Dhanisiri Project      & 83.37                \\
Sukla Irr. Project     & 27.4                 \\
Jamuna Irr, Project    & 41.01                \\
Champamati Irr.Project & 24.99               \\
\hline
\end{tabular}
\caption{Irrigation Potential for different Projects in network}
\label{Table:1}
\end{table}

\par 
\noindent
You can clearly see from the network that it has 17 total nodes. Next we calculate the Clustering Coefficient and the Link Density for the Brahmaputra river basin network, which comes out to be0 and 0.058. represents how densely the nodes of the network are connected to each other. High Link Density means a highly connected network. The Elevation Gradient, which is a topographical characteristic of the network, is around 5950 meters.

\section{STRUCTURAL PROPERTIES}
\noindent
Each network here is modeled as a graph G(V,E) where V denotes the number of vertices or nodes in the graph and E denotes the number of edges connecting the two particular vertices. The two main properties of the network that have been studied here is the link density, and the clustering coefficient of the network. This has been summarized in Table \ref{Table:1} below. The reason for that is the practical reality of the system of interest here, i.e. the water distribution system.
\par
\noindent
The Clustering Coefficient of a node is the probability that any two neighbours of a
node are also neighbours of each other\cite{10.1093/oso/9780198821939.001.0001}. The Clustering Coefficient of the entire system is the average of Clustering Coefficients of nodes in the system. Non-zero clustering coefficients are not expected in such systems as it does not make logical sense in a civil engineering or in a planning and management perspective. We do not have non-zero clustering coefficients also because what it would geographically mean is three barrages or dams connected in a triangular manner, which implies two possibilities in terms of river connectivity. The first being, three different rivers intersecting and we have a dam or barrage at each of the intersection points to control the water flow or for storage purposes. The second possibility is, bifurcation of a particular river in two separate branches at a particular point and later recombination of these two branches again at a particular spatially separated location. Now if we have a dam or barrage constructed at the bifurcation and recombination location, along with a dam/barrage in one of the two bifurcated branches, then we would have a non-zero clustering coefficient in such systems. But empirically we do not observe such types of water dam construction projects and hence we do not expect a non-zero clustering coefficient for our 15 river basin networks; this is also clearly reflected in the table \ref{Table:1} below. 
\par
\noindent
The other property that has been of particular importance is the Link Density (LD) of the river basin Networks. The Link Density (LD) is generally defined for an undirected network as
\begin{equation}
   LD = \frac{2*m}{n*(n-1)} 
\end{equation}
here $m$ is the number of edges in the Graph and $n$ are the total number of vertices in the graph. For once, we put aside the directedness of our network and study the Link density for our networks.The reason we are interested in studying the Link density is because a relatively higher Link Density would mean a densely connected network. And in such networks, the spread of information is easier and faster as compared to the network with lower Link Density. And the reason it matters in the context of Water Dam Networks is because it gives us information about which river basin network is more connected, which can lead to more uniform and situation-wise distribution of water load. But it also carries a disadvantage in terms of Cascading failures in these networks. A more detailed review of cascading failures in networks is given here \cite{10.1093/comnet/cnaa013}.  The more densely connected the network is, the easier it gets for the failure to propagate through the whole network. Particularly in Water Dam Networks, this might represent a case where due to cloud burst or any civil or mechanical failure, a certain dam or barrage loses its ability to function fully or partially and this may lead to adverse effects on the neighboring nodes (in this case a dam or a barrage). If the network is densely connected i.e. having a high Link Density then there is a chance that this failure propagates throughout the network, and in this case the consequences are adverse. The Link Densities (LD) for all the 15 river basins have been summarized in the table \ref{Table:2} below.
\begin{table}[h]
\begin{tabular}{|l|c|c|}
\hline
Basin        & Link Density(LD) & Clustering Coefficient \\
\hline
Brahmani     & 0.1              &         0              \\
Barak        & 0.166            & 0                      \\
Brahmaputra  & 0.058            & 0                      \\
Cauvery      & 0.1              & 0                      \\
Godavri      & 0.3334           & 0                      \\
Indus        & 0.0714           & 0                      \\
Krishna      & 0.0667           & 0                      \\
Mahanadi     & 0.0834           & 0                      \\
Mahi         & 0.125            & 0                      \\
Narmada      & 0.071            & 0                      \\
Ganga        & 0.025            & 0                      \\
Pennar       & 0.0834           & 0                      \\
Sabarmati    & 0.1              & 0                      \\
Subarnarekha & 0.0834           & 0                      \\
Tapi         & 0.091            & 0                     \\
\hline
\end{tabular}
\caption{LD and Clustering Coefficients.}
\label{Table:2}
\end{table}

\par
\noindent
Apart from Link Density and the Clustering Coefficient of the network, we have also calculated the total number of nodes for each river basin network. We have also calculated the total nodes with degree 1, 2, 3 ,and, 4 and above, for each river basin network. This information is summarized in the table \ref{Table:3} below.

\begin{table}[!h]
\begin{tabular}{|l|c|c|c|c|c|}
\hline
Basin        & d(1) & d(2) & d(3) & d($>=4$) & Total \\
\hline
Brahmani     & 3    & 6    & 1    & 0      & 10    \\
Barak        & 4    & 1    & 0    & 1      & 6     \\
Brahmaputra  & 13   & 2    & 1    & 1      & 17    \\
Cauvery      & 6    & 0    & 4    & 0      & 10    \\
Godavri      & 16   & 9    & 1    & 4      & 30    \\
Indus        & 4    & 8    & 2    & 0      & 14    \\
Krishna      & 7    & 3    & 5    & 0      & 15    \\
Mahanadi     & 6    & 4    & 1    & 1      & 12    \\
Mahi         & 4    & 3    & 0    & 1      & 8     \\
Narmada      & 8    & 4    & 0    & 2      & 14    \\
Ganga        & 18   & 18   & 2    & 2      & 40    \\
Pennar       & 5    & 5    & 1    & 1      & 12    \\
Sabarmati    & 7    & 0    & 2    & 1      & 10    \\
Subarnarekha & 6    & 4    & 1    & 1      & 12    \\
Tapi         & 7    & 1    & 1    & 2      & 11    \\
\hline
\end{tabular}
\caption{d(x) = number of nodes with degree x.}
\label{Table:3}
\end{table}

\par
\noindent
We have also tried to see if there is any correlation between the Link Density and the total number of nodes. You can see the Figure \ref{fig:2} below.
\begin{figure}[h]
  \includegraphics[width=\linewidth]{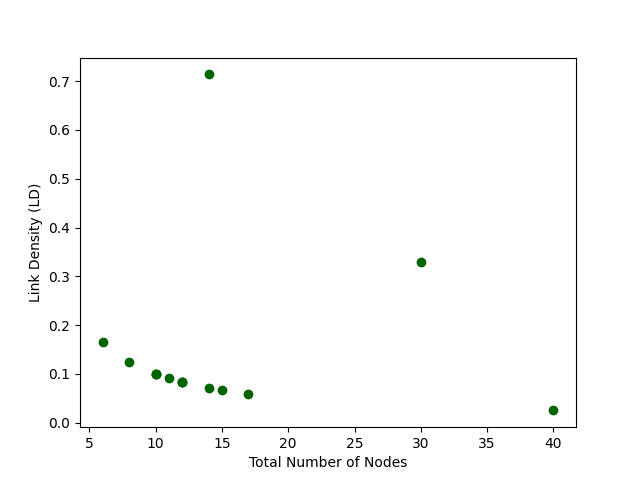}
  \caption{Link Density vs Total Nodes in each network.}
  \label{fig:2}
\end{figure}
\noindent
The correlation for the above plot has been calculated using the numpy module in python which gives the correlation coefficient to be 0.024. This is positive but a very weak correlation so not much can be predicted using this relation.

\section{IRRIGATION POTENTIAL}
\noindent
One of the reasons why the Water Distribution Networks like the networks of Dams and Barrages is important for agrarian countries like India is that the majority of the population in rural areas is occupied by agricultural activities for whom water supply is of vital importance. Thus the amount of irrigation potential these river basin networks contain, greatly influences the country's economy and the agricultural output. The Irrigation potential for each river basin network is measured in units of Thousand Hectare (Th Ha) and is summarized in table \ref{Table:4} below \cite{India-WRIS.2012}. 

\begin{table}[h]
\begin{tabular}{|l|c|}
\hline
Basin        & Irrigation Potential \\
\hline
Brahmani     & 670.67               \\
Barak        & 27.72                \\
Brahmaputra  & 703.77               \\
Cauvery      & 1403.66              \\
Godavri      & 2760.82              \\
Indus        & 6598.933             \\
Krishna      & 4212.15              \\
Mahanadi     & 1808.98              \\
Mahi         & 466.249              \\
Narmada      & 3568.3               \\
Ganga        & 14056.56             \\
Pennar       & 263.84               \\
Sabarmati    & 226.85               \\
Subarnarekha & 581.222              \\
Tapi         & 514.25               \\
\hline
\end{tabular}
\caption{Irrigation Potential (Thousand Hectares)}
\label{Table:4}
\end{table}

\par
\begin{figure}[ht]
\begin{subfigure}{.5\textwidth}
  \centering
  % include first image
  \includegraphics[width=.8\linewidth]{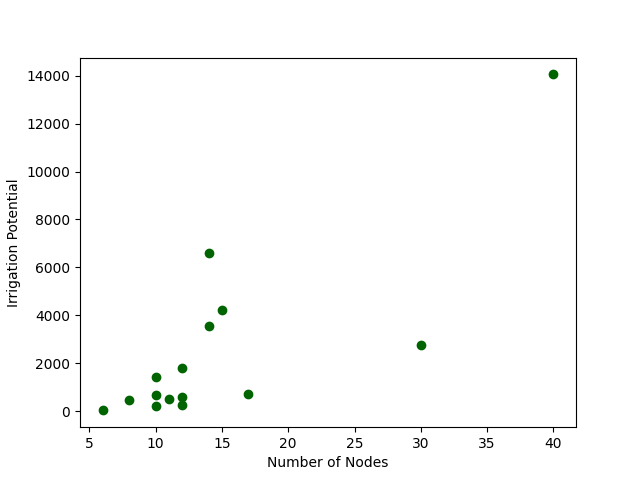}  
  \caption{Relation between Irrigation Potential and Total Nodes }
  \label{fig:3a}
\end{subfigure}
\begin{subfigure}{.5\textwidth}
  \centering
  % include second image
  \includegraphics[width=.8\linewidth]{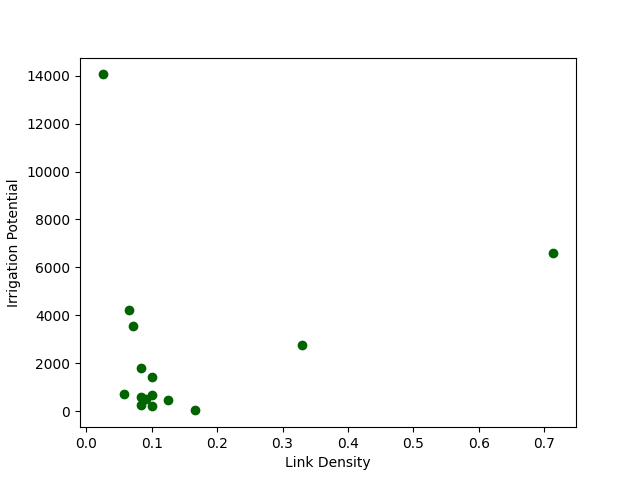}  
  \caption{Relation between Irrigation Potential and Link Density}
  \label{fig:3b}
\end{subfigure}
\caption{Relation of Irrigation Potential with Link Density and total nodes.}
\label{fig:3}
\end{figure}

\noindent
Here, I have tried to see if there is any correlation between Irrigation potential and Link Density (LD), and Irrigation Potential and Number of Nodes in Network, for each river basin. 
There is a very weak correlation between Irrigation Potential and Link Density(LD) of the networks. But there seems to be a good correlation between Irrigation Potential and Number of Nodes in each network of different river basins. This can be seen in the figure \ref{fig:3}.

\par
\noindent
To quantify our results, the correlation test has been performed. The correlation coefficients have been calculated for both Figure \ref{fig:3a} and Figure \ref{fig:3b}. The correlation coefficient ($r$) is given by :
\begin{equation}
    r=\frac{\Sigma(x_{i}-\bar{x})(y_{i}-\bar{y})}{\sqrt{\Sigma(x_{i}-\bar{x})^{2}(y_{i}-\bar{y})^2}}
\end{equation}
where $x$ and $y$ are two sets such that $x_{i} \in x$ and $y_{i}\in y$, and $\bar{x}$ and $\bar{y}$ represents the mean of two sets $x$ and $y$, respectively.

\par
\noindent
The Correlation coefficient between Irrigation Potential and Link Density comes out to be 0.176, approximately, which shows a very weak correlation. But the correlation coefficient between Irrigation Potential and Number of nodes of the network comes out to be 0.805, approximately. This shows a very strong correlation.

\subsection{Mathematical Model}
To find a mathematical model for this correlation, we assume $y = cx^{a}$ relation as an ansatz. And using the linregress module from scipy in Python, we estimate the parameters c and a for our data. Here y represents the Irrigation Potential and x represents the total number of nodes in the network. When we take the logarithm on both the sides we get:
\begin{equation}
    \log(y) = \log(c) + a\log(x)
\end{equation}

\par
\noindent

\begin{figure}[h]
  \includegraphics[width=\linewidth]{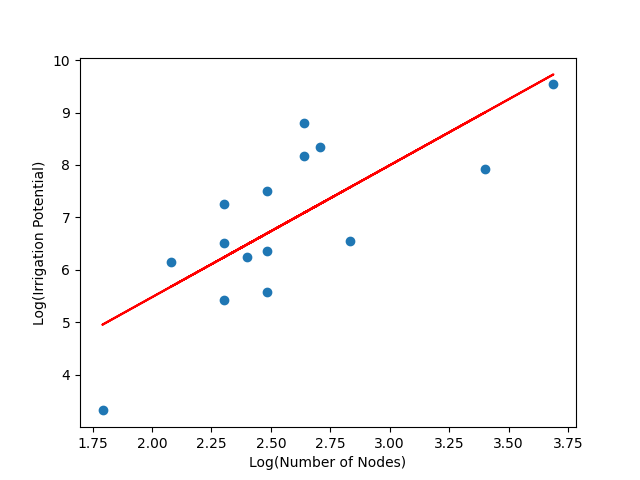}
  \caption{Logarithmic Graph for Irrigation Potential and Nodes.}
  \label{fig:4}
\end{figure}

Figure \ref{fig:4} shows the Logarithmic graph for Irrigation Potential and the total Nodes.

\par
\noindent
From the Scipy module we import the linregress and we estimate the parameters. This gives us $a= 2.514$, $\log(c)= 0.452$, $r=0.764$, Standard Error=0.589, and, P-value= 0.00092.

\par
\noindent
This positive correlation between the Irrigation potential and the Number of Nodes simply signifies that if we increase the number of dams/barrages, we will have higher irrigation potential and cultivable command over a larger area of land which is expected. But what is also of importance is the estimated mathematical relationship which has been put forth here. This mathematical relation can be used to get an approximate estimate of how much increase in the irrigation potential can be expected in the region on addition of certain number of dams in the area. Ofcourse, the geographical reasons here do play a role in deciding this, but it still gives a logical estimate over which future work can be done to include additional factors and make this estimate further precise.

\section{TOPOGRAPHIC RESPONSE}
\noindent
Here I have investigated the presence of any correlation between the Elevation Gradient of the Basin measured in meters and the Link Density of that river basin network. Correlation between Elevation Gradient and the Total number of nodes has also been investigated. The data related to the Elevation Gradient of each basin has been summarized below in table \ref{Table:5} 

\begin{table}[h]
\begin{tabular}{|l|c|}
\hline
Basin        & Elevation Gradient \\
\hline
Brahmani     & 1500               \\
Barak        & 3000               \\
Brahmaputra  & 5950               \\
Cauvery      & 3000               \\
Godavri      & 1500               \\
Indus        & 5900               \\
Krishna      & 2000               \\
Mahanadi     & 1500               \\
Mahi         & 1000               \\
Narmada      & 1500               \\
Ganga        & 6000               \\
Pennar       & 1500               \\
Sabarmati    & 1500               \\
Subarnarekha & 1500               \\
Tapi         & 1500               \\
\hline
\end{tabular}
\caption{Approximate Elevation Gradient in meters}
\label{Table:5}
\end{table}

\begin{figure}[ht]
\begin{subfigure}{.5\textwidth}
  \centering
  % include first image
  \includegraphics[width=.8\linewidth]{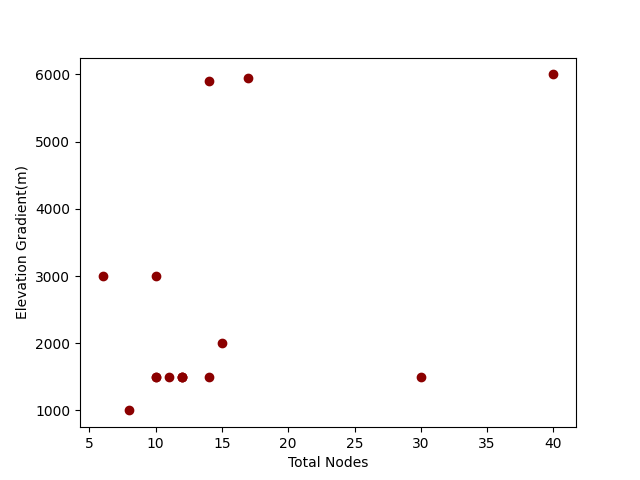}  
  \caption{Relation between Elevation Gradients and Total Nodes }
  \label{fig:5a}
\end{subfigure}
\begin{subfigure}{.5\textwidth}
  \centering
  % include second image
  \includegraphics[width=.8\linewidth]{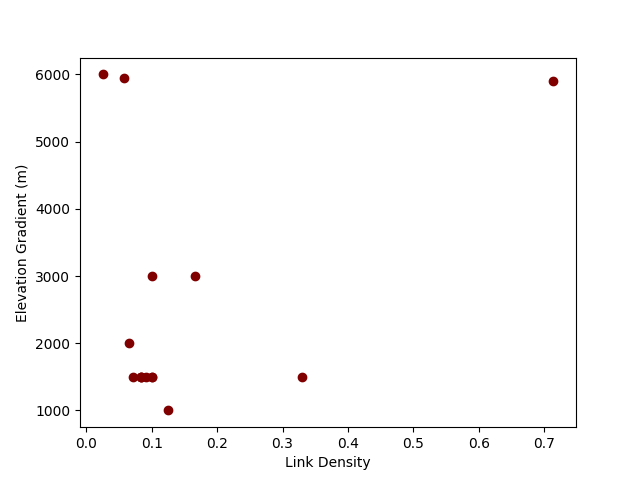}  
  \caption{Relation between Elevation Gradient and Link Density}
  \label{fig:5b}
\end{subfigure}
\begin{subfigure}{.5\textwidth}
  \centering
  % include third image
  \includegraphics[width=.8\linewidth]{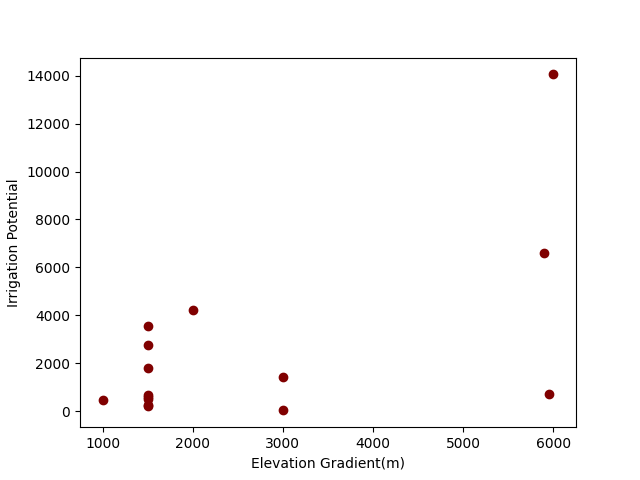}  
  \caption{Relation between Elevation Gradient and Irrigation Potential}
  \label{fig:5c}
\end{subfigure}
\caption{Relation of Elevation Gradient with Link Density, total nodes and Irrigation Potential.}
\label{fig:5}
\end{figure}

The plots for the same can be found in figure \ref{fig:5}.

\par
\noindent
The Coefficient of Correlation for the figure \label{fig:5a} which is the Elevation Gradient vs the Total nodes of the network graph, comes out to be 0.458, approximately. And for figure \label{fig:5b} i.e. Elevation Gradient vs the Link Density (LD) comes out to be 0.340, approximately. This tells us that there is no strong correlation between the Elevation Gradient and Link Density of the network or the total number of nodes in the network. This also tells us that the response of the water dam networks to the topography of the river basin is weak and the salient features like the Elevation Gradient of the region are not reciprocated in planning the water distribution systems.

\par 
\noindent
The Coefficient of Correlation for the figure \label{fig:5c} turns out to be 0.617, approximately. This is a decent correlation between the Elevation Gradient and the Irrigation Potential of the networks, implying that the networks with higher topographical elevation gradient also have better Irrigation Potential. This also explains the outlier that we see in figure \ref{fig:3b}, where even though the Link Density of the Ganga river basin is very low, the Irrigation Potential is above 14000 Th Ha since the Elevation Gradient of the Ganga River Basin is very high (around 6000 meters). From our analysis, we have seen that the Irrigation Potential and the total number of nodes in the network show a strong correlation. And we also know that the Elevation Gradient and the total number of nodes in the network also show a moderate correlation. So we can say that total number of nodes, the Elevation Gradient and the Irrigation potential of the networks are positively correlated with each other.

\section{DISCUSSIONS AND CONCLUSION}
\noindent
In this paper, I have taken a complex network approach to understand the salient features of Indian Water Dam Networks. We represent water dams and barrages as nodes and the connection between them with directed edges. So it's a directed network, since the river flow has a certain fixed direction. We have tried to see the correlation between various structural properties of the network, and between topographical properties and the structural properties of the network for each of the 15 river basins. This has helped us in understanding the stability, topographic response and Irrigation Potential for the networks of each river basin. The observations have been summarized here :

\begin{enumerate}
    \item In the paper it is observed that the Link Density of a network of a particular river basin is not correlated with the number of nodes, i.e. Dams or Barrages in the same river basin network.
    \item Irrigation potential  for each of the river basins have been listed in the paper. Then we investigated the possibility of correlation between Irrigation Potential and Link Density, which eventually turned out to be very weak. We also investigated the correlation between irrigation potential and total number of nodes in the network, which in accordance with what is logically expected, did show a strong positive correlation. We have tried to quantify our results by measuring the correlation coefficients for each case. We have also presented a mathematical model, more appropriately an ansatz, for the relation between Irrigation potential and total number of nodes, which would help estimate the relative increase in irrigation potential for a region for a certain number of increase in the number of dams of barrages, given certain number of dams and Irrigation potential of a the time. Improving on this mathematical ansatz in future work can lead to more accurate predictions as well.
    \item In the last section of the paper, the response of the structural features of the network to the topography of the region has been tested. The Correlation between the Elevation Gradient and the number of nodes in the network shows a weak to moderate correlation, whose implications can be explained more rigorously both in quantitative and qualitative manner by doing detailed research in future work. The Correlation between the Elevation Gradient and the Link Density (LD) shows a weak positive correlation as well. So overall we can conclude that the response of the river basin network to the topography of that region is not strong enough. And whether better planning in terms of improving this topographical response of the network leads to some practical economic, social or agricultural benefits is subject to more detailed future work. This is a point of crucial importance, since more efficient planning can be done while designing such water management systems or programs by doing the network analysis for it.
    \item In the last section we also investigate the correlation between the elevation gradient and the irrigation potential of the networks, which shows a decently positive correlation between them. From our previous analysis we can conclude that the irrigation potential, elevation gradient and the total number of nodes of the network are positively correlated with each other. This gives a better understanding to the strategists who are planning such irrigation projects on a varied topographical layout of a region to maximize the irrigational output and also building optimum amount of dams in the region.
\end{enumerate}

\par
\noindent
In summary, we can say that studying these structural properties of these systems like the Water Dam Networks can help us understand the system much more quantitatively. This can help the future policy makers in making more informed quantitative decisions in planning these water management projects. Complex Networks approach is a very elegant, simplistic and rigorous method in understanding such interacting many component systems. In no way can we say that these water dam networks are entirely understood for India, and in that regard, for any other countries as well. Future work can build up on this to make the model more and more efficient in terms of predictions and fundamental understanding of the complexity, stability, vulnerability and efficiency of these water dam networks. This would lead to a good advantage to avoid any future risks related to failure propagation in these systems which in this case carry adverse consequences like floods and droughts. A good understanding of this system can also help in making decisions that may result in increase of agricultural productivity and implicitly impact the economic output of the country positively. Improved planning in these regards can also contribute to increasing the standard of living for the people in rural areas involved in agrarian activities. Since, it would result in more systematic water supply and their agricultural productivity will increase. Hence, future work in understanding these water dam networks through a complex network approach is absolutely essential and this lays down a basic groundwork over which much more detailed analysis can be done.
\vspace{0.6cm}
\begin{acknowledgments}

\par
\noindent
I would like to express my sincere gratitude to the Central Water Commission (CWC), Ministry of Water Resources, Government of India and the Indian Space Research Organisation (ISRO) for the valuable information they have compiled in the “ River Basin Atlas of India” which has been extensively used in the paper. I would also like to thank Prof. Arijit Bhattacharyay, Department of Physics, Indian Institute of Science Education and Research (IISER), Pune, India, for giving valuable suggestions whenever and wherever I approached him regarding the topic of interest.
\end{acknowledgments}

\appendix

%\section{Appendixes}

% The \nocite command causes all entries in a bibliography to be printed out
% whether or not they are actually referenced in the text. This is appropriate
% for the sample file to show the different styles of references, but authors
% most likely will not want to use it.
\nocite{*}

\bibliographystyle{plain}
\bibliography{apssamp}% Produces the bibliography via BibTeX.

\providecommand{\noopsort}[1]{}\providecommand{\singleletter}[1]{#1}%
\begin{thebibliography}{10}

\bibitem{RevModPhys.74.47}
R\'eka Albert and Albert-L\'aszl\'o Barab\'asi.
\newblock Statistical mechanics of complex networks.
\newblock {\em Rev. Mod. Phys.}, 74:47--97, Jan 2002.

\bibitem{Albert2000}
R{\'e}ka Albert, Hawoong Jeong, and Albert-L{\'a}szl{\'o} Barab{\'a}si.
\newblock Error and attack tolerance of complex networks.
\newblock {\em Nature}, 406(6794):378--382, Jul 2000.

\bibitem{9521783}
Ali~Moradi Amani and Mahdi Jalili.
\newblock Power grids as complex networks: Resilience and reliability analysis.
\newblock {\em IEEE Access}, 9:119010--119031, 2021.

\bibitem{doi:10.1126/science.286.5439.509}
Albert-László Barabási and Réka Albert.
\newblock Emergence of scaling in random networks.
\newblock {\em Science}, 286(5439):509--512, 1999.

\bibitem{BOCCALETTI2006175}
S.~Boccaletti, V.~Latora, Y.~Moreno, M.~Chavez, and D.-U. Hwang.
\newblock Complex networks: Structure and dynamics.
\newblock {\em Physics Reports}, 424(4):175--308, 2006.

\bibitem{Buhl2006}
J.~Buhl, J.~Gautrais, N.~Reeves, R.~V. Sol{\'e}, S.~Valverde, P.~Kuntz, and G.~Theraulaz.
\newblock Topological patterns in street networks ofself-organized urban settlements.
\newblock {\em The European Physical Journal B - Condensed Matter and Complex Systems}, 49(4):513--522, Feb 2006.

\bibitem{PhysRevE.73.066107}
Alessio Cardillo, Salvatore Scellato, Vito Latora, and Sergio Porta.
\newblock Structural properties of planar graphs of urban street patterns.
\newblock {\em Phys. Rev. E}, 73:066107, Jun 2006.

\bibitem{India-WRIS.2012}
Indian Space Researc~Organisation Central Water~Commission.
\newblock River basin atlas of india.
\newblock 2012.

\bibitem{PhysRevLett.86.3682}
Reuven Cohen, Keren Erez, Daniel ben Avraham, and Shlomo Havlin.
\newblock Breakdown of the internet under intentional attack.
\newblock {\em Phys. Rev. Lett.}, 86:3682--3685, Apr 2001.

\bibitem{CRUCITTI200492}
Paolo Crucitti, Vito Latora, and Massimo Marchiori.
\newblock A topological analysis of the italian electric power grid.
\newblock {\em Physica A: Statistical Mechanics and its Applications}, 338(1):92--97, 2004.
\newblock Proceedings of the conference A Nonlinear World: the Real World, 2nd International Conference on Frontier Science.

\bibitem{article}
Luciano da~F.~Costa, F~Costa, Francisco Rodrigues, and Alexandre Cristino.
\newblock Complex networks: The key to systems biology.
\newblock {\em Genetics and Molecular Biology}, 31, 12 2007.

\bibitem{doi:10.1080/00018730110112519}
S.~N. Dorogovtsev and J.~F.~F. Mendes.
\newblock Evolution of networks.
\newblock {\em Advances in Physics}, 51(4):1079--1187, 2002.

\bibitem{FORSBERG2023129072}
Samuel Forsberg, Karin Thomas, and Mikael Bergkvist.
\newblock Power grid vulnerability analysis using complex network theory: A topological study of the nordic transmission grid.
\newblock {\em Physica A: Statistical Mechanics and its Applications}, 626:129072, 2023.

\bibitem{10.5120/ijca2015905952}
Sanjeev Kumar~Yadav Hradesh~Kumar.
\newblock Investigating social network as complex network and dynamics of user activities.
\newblock {\em International Journal of Computer Applications}, 125(7):13--18, September 2015.

\bibitem{Masucci2009}
A.~P. Masucci, D.~Smith, A.~Crooks, and M.~Batty.
\newblock Random planar graphs and the london street network.
\newblock {\em The European Physical Journal B}, 71(2):259--271, Sep 2009.

\bibitem{doi:10.1137/S003614450342480}
M.~E.~J. Newman.
\newblock The structure and function of complex networks.
\newblock {\em SIAM Review}, 45(2):167--256, 2003.

\bibitem{10.1093/acprof:oso/9780199206650.001.0001}
Mark Newman.
\newblock {\em {Networks: An Introduction}}.
\newblock Oxford University Press, 03 2010.

\bibitem{doi:10.1061/(ASCE)0733-9496(2005)131:1(58)}
Avi Ostfeld.
\newblock Water distribution systems connectivity analysis.
\newblock {\em Journal of Water Resources Planning and Management}, 131(1):58--66, 2005.

\bibitem{Sanhedrai2022}
Hillel Sanhedrai, Jianxi Gao, Amir Bashan, Moshe Schwartz, Shlomo Havlin, and Baruch Barzel.
\newblock Reviving a failed network through microscopic interventions.
\newblock {\em Nature Physics}, 18(3):338--349, Mar 2022.

\bibitem{Sanhedrai2023}
Hillel Sanhedrai and Shlomo Havlin.
\newblock Sustaining a network by controlling a fraction of nodes.
\newblock {\em Communications Physics}, 6(1):22, Jan 2023.

\bibitem{SITZENFREI2021117359}
Robert Sitzenfrei.
\newblock Using complex network analysis for water quality assessment in large water distribution systems.
\newblock {\em Water Research}, 201:117359, 2021.

\bibitem{https://doi.org/10.1029/2020WR027929}
Robert Sitzenfrei, Qi~Wang, Zoran Kapelan, and Dragan Savić.
\newblock Using complex network analysis for optimization of water distribution networks.
\newblock {\em Water Resources Research}, 56(8):e2020WR027929, 2020.
\newblock e2020WR027929 2020WR027929.

\bibitem{Smolyak2020}
Alex Smolyak, Orr Levy, Irena Vodenska, Sergey Buldyrev, and Shlomo Havlin.
\newblock Mitigation of cascading failures in complex networks.
\newblock {\em Scientific Reports}, 10(1):16124, Sep 2020.

\bibitem{w15081621}
Federico Spizzo, Giovanni Venaruzzo, Matteo Nicolini, and Daniele Goi.
\newblock Water distribution network partitioning based on complex network theory: The udine case study.
\newblock {\em Water}, 15(8), 2023.

\bibitem{Tamhane_2023}
Vedang Tamhane and G~Ambika.
\newblock Structure and stability of the indian power transmission network.
\newblock {\em Journal of Physics: Complexity}, 4(2):025014, may 2023.

\bibitem{10.1093/oso/9780198821939.001.0001}
Stefan Thurner, Peter Klimek, and Rudolf Hanel.
\newblock {\em {Introduction to the Theory of Complex Systems}}.
\newblock Oxford University Press, 09 2018.

\bibitem{10.1093/comnet/cnaa013}
Lucas~D Valdez, Louis Shekhtman, Cristian~E La~Rocca, Xin Zhang, Sergey~V Buldyrev, Paul~A Trunfio, Lidia~A Braunstein, and Shlomo Havlin.
\newblock {Cascading failures in complex networks}.
\newblock {\em Journal of Complex Networks}, 8(2):cnaa013, 05 2020.

\bibitem{Watts1998}
Duncan~J. Watts and Steven~H. Strogatz.
\newblock Collective dynamics of `small-world' networks.
\newblock {\em Nature}, 393(6684):440--442, Jun 1998.

\bibitem{10.1063/1.3540339}
Alireza Yazdani and Paul Jeffrey.
\newblock {Complex network analysis of water distribution systems}.
\newblock {\em Chaos: An Interdisciplinary Journal of Nonlinear Science}, 21(1):016111, 03 2011.

\end{thebibliography}

\newpage
\section{River Basin Networks}
Here I have added the Networks for each of the 15 river Basin Networks. The node size are made to be proportional to their respective degrees and the arrows show the link direction between nodes.
\begin{figure}[h]
  \includegraphics[width=\linewidth]{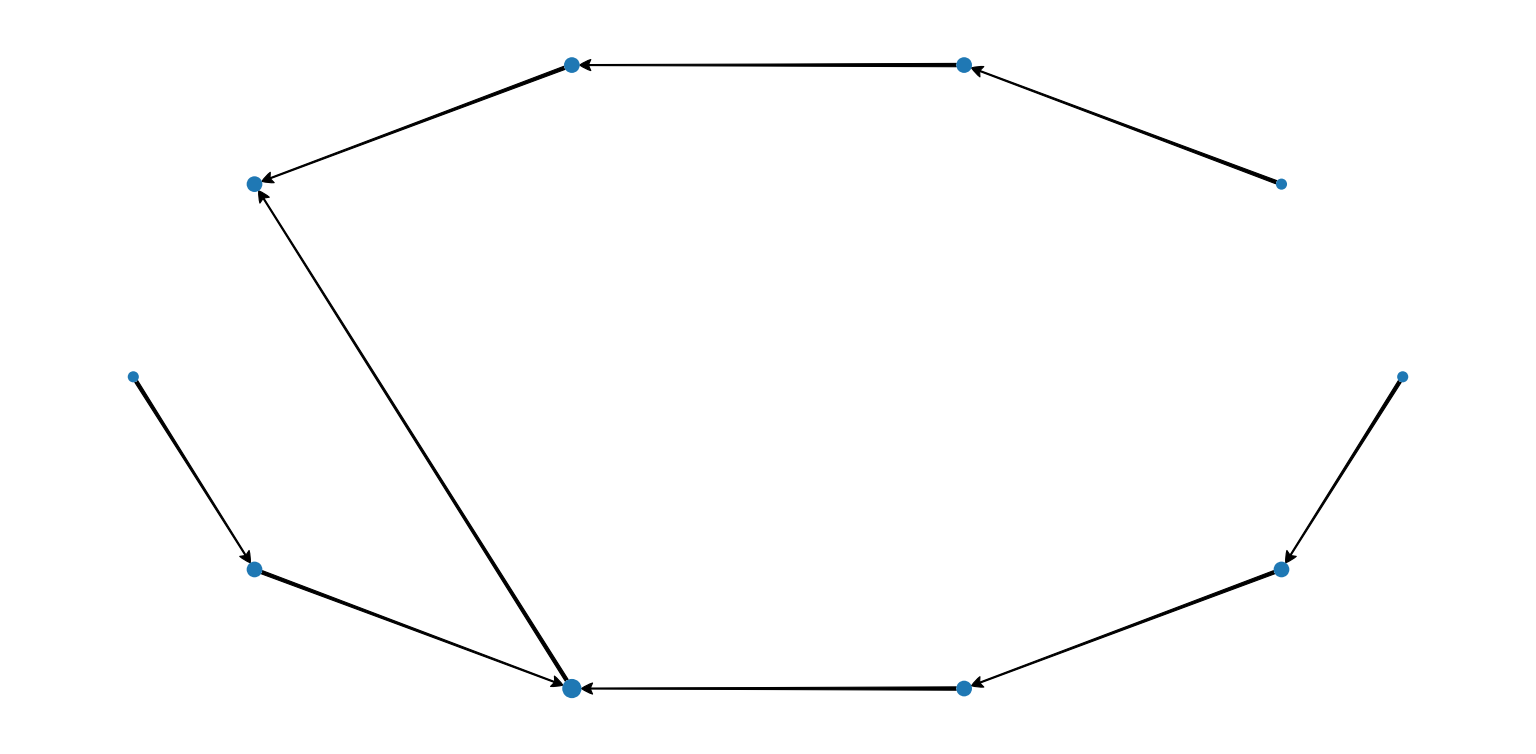}
  \caption{Brahmani and Baitarani Basin.}
\end{figure}

\begin{figure}[h]
  \includegraphics[width=\linewidth]{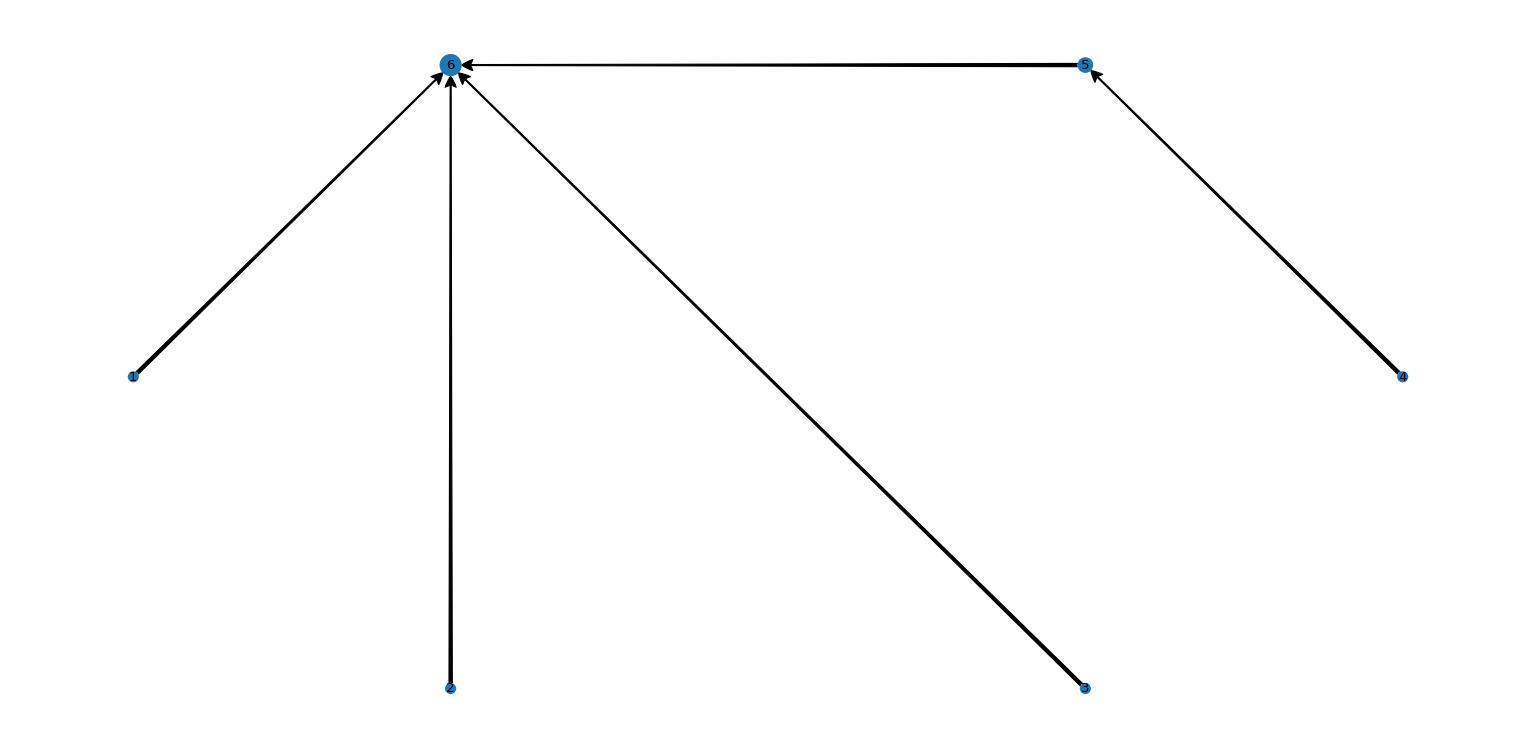}
  \caption{Barak Basin.}
\end{figure}

\begin{figure}[h]
  \includegraphics[width=\linewidth]{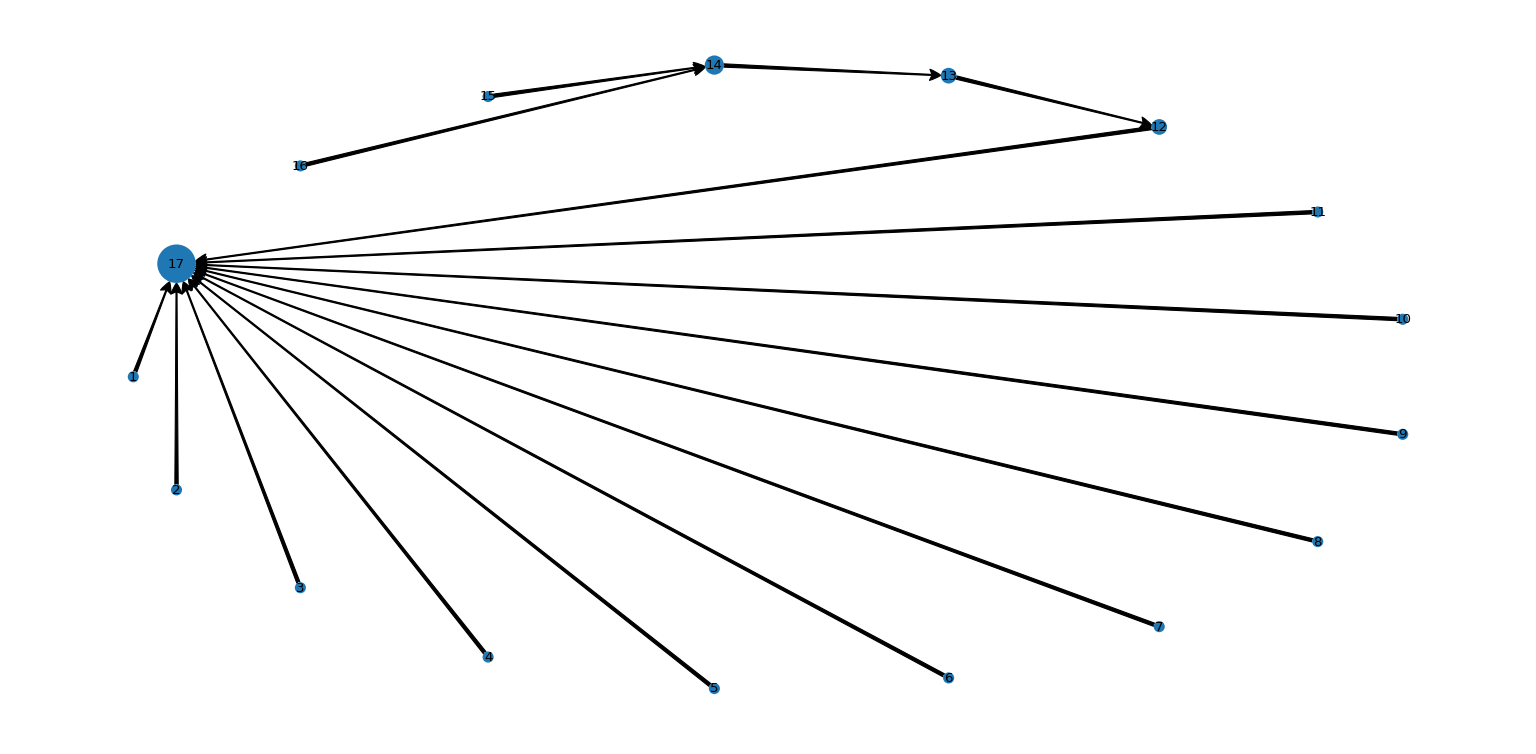}
  \caption{Brahmaputra Basin.}
\end{figure}

\begin{figure}[h]
  \includegraphics[width=\linewidth]{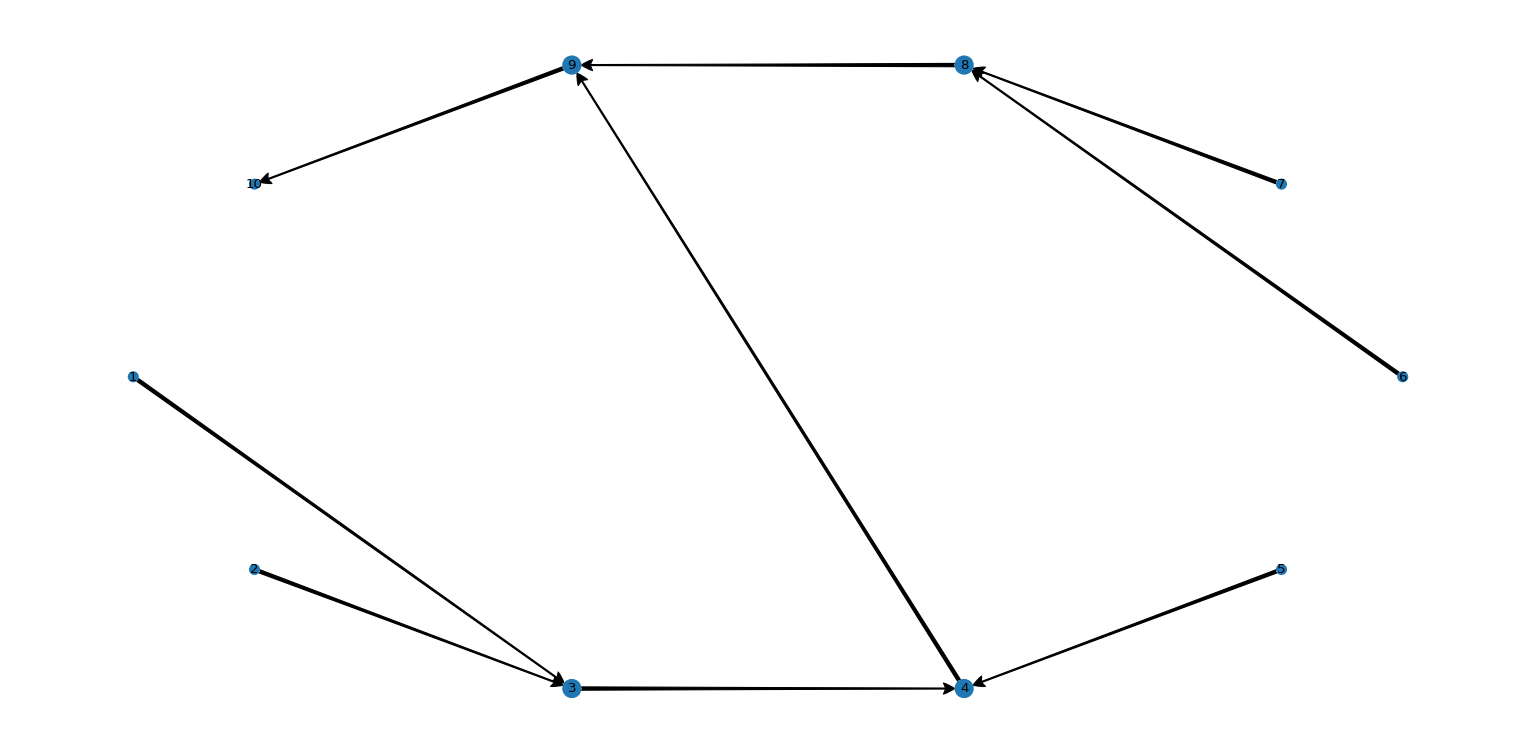}
  \caption{Cauvery Basin.}
\end{figure}

\begin{figure}[h]
  \includegraphics[width=\linewidth]{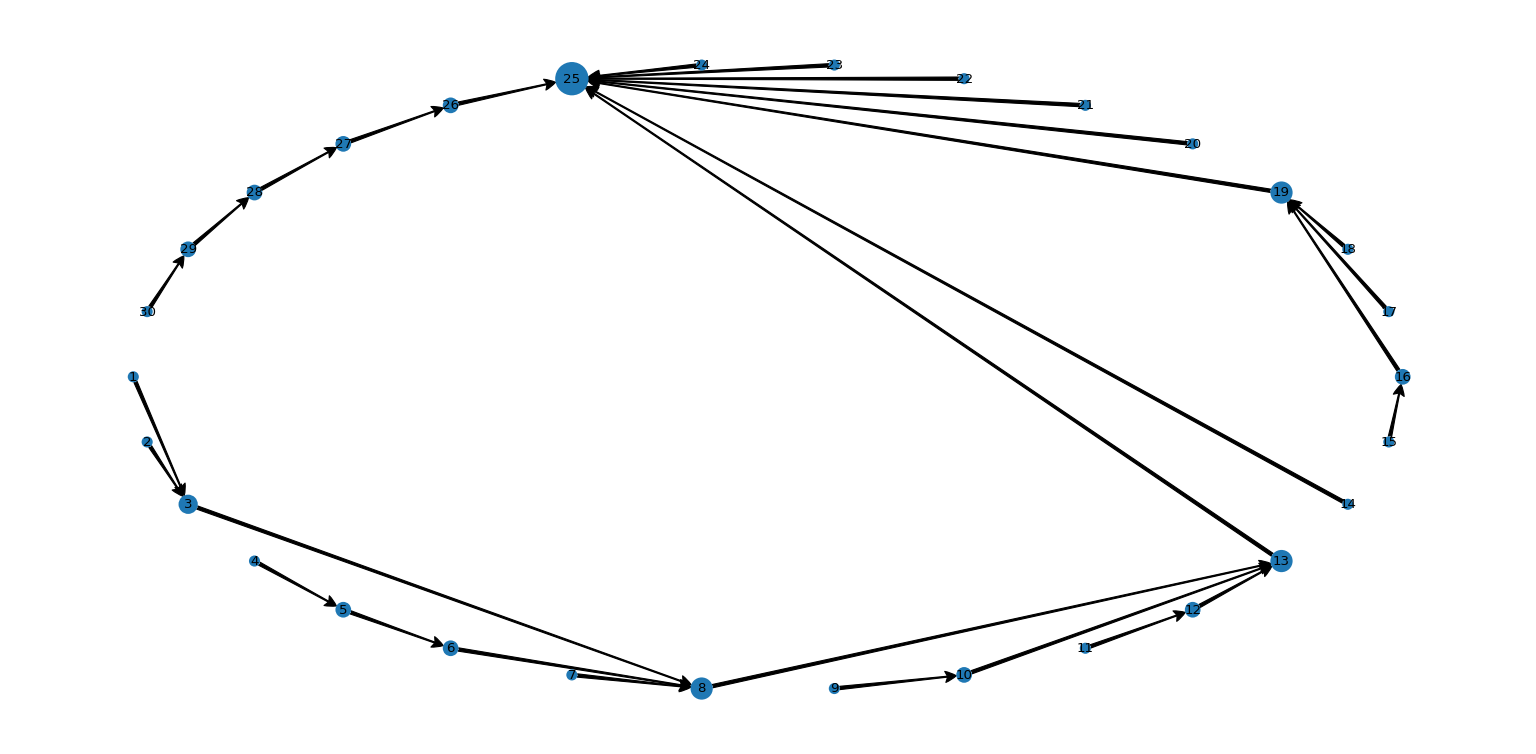}
  \caption{Godavri Basin.}
\end{figure}

\begin{figure}[h]
  \includegraphics[width=\linewidth]{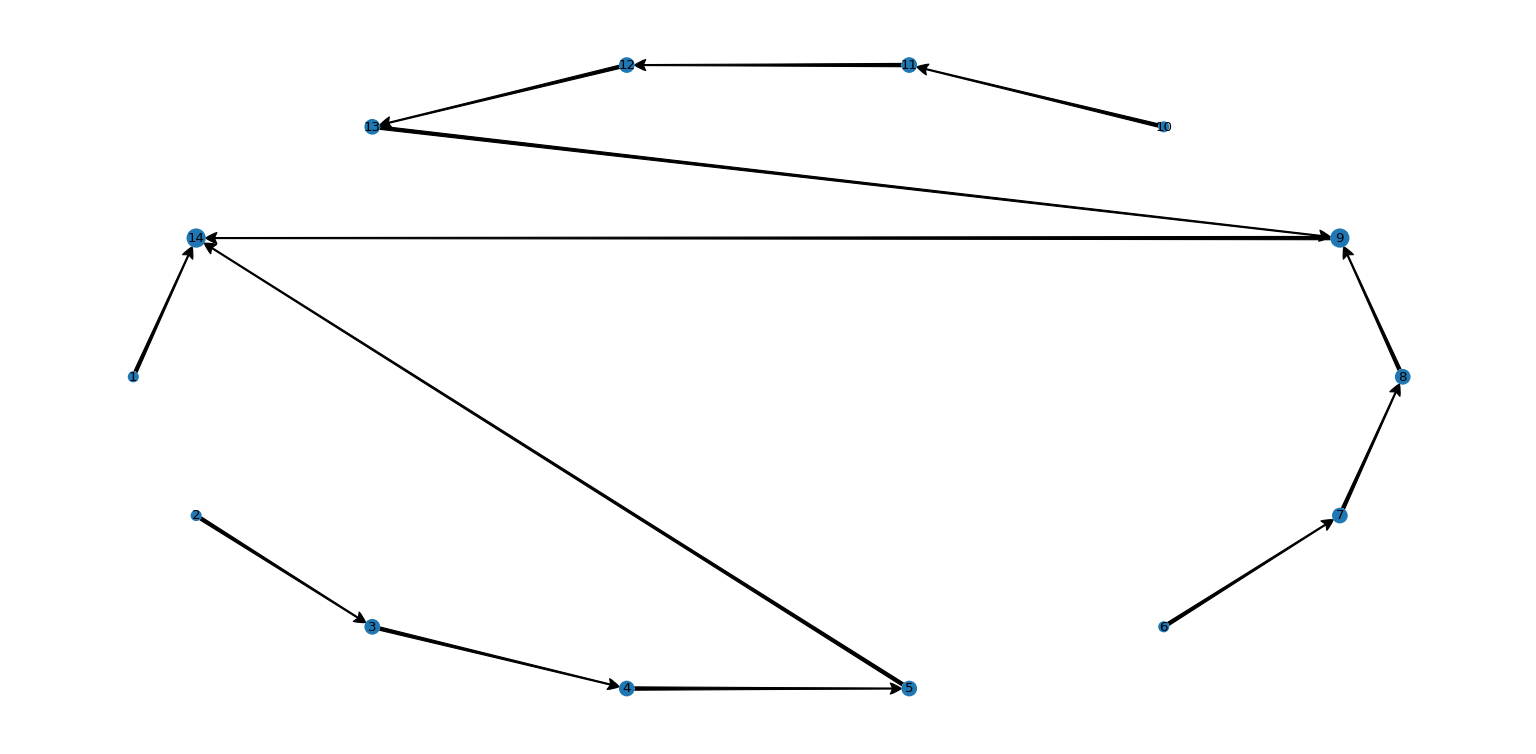}
  \caption{Indus Basin.}
\end{figure}

\begin{figure}[h]
  \includegraphics[width=\linewidth]{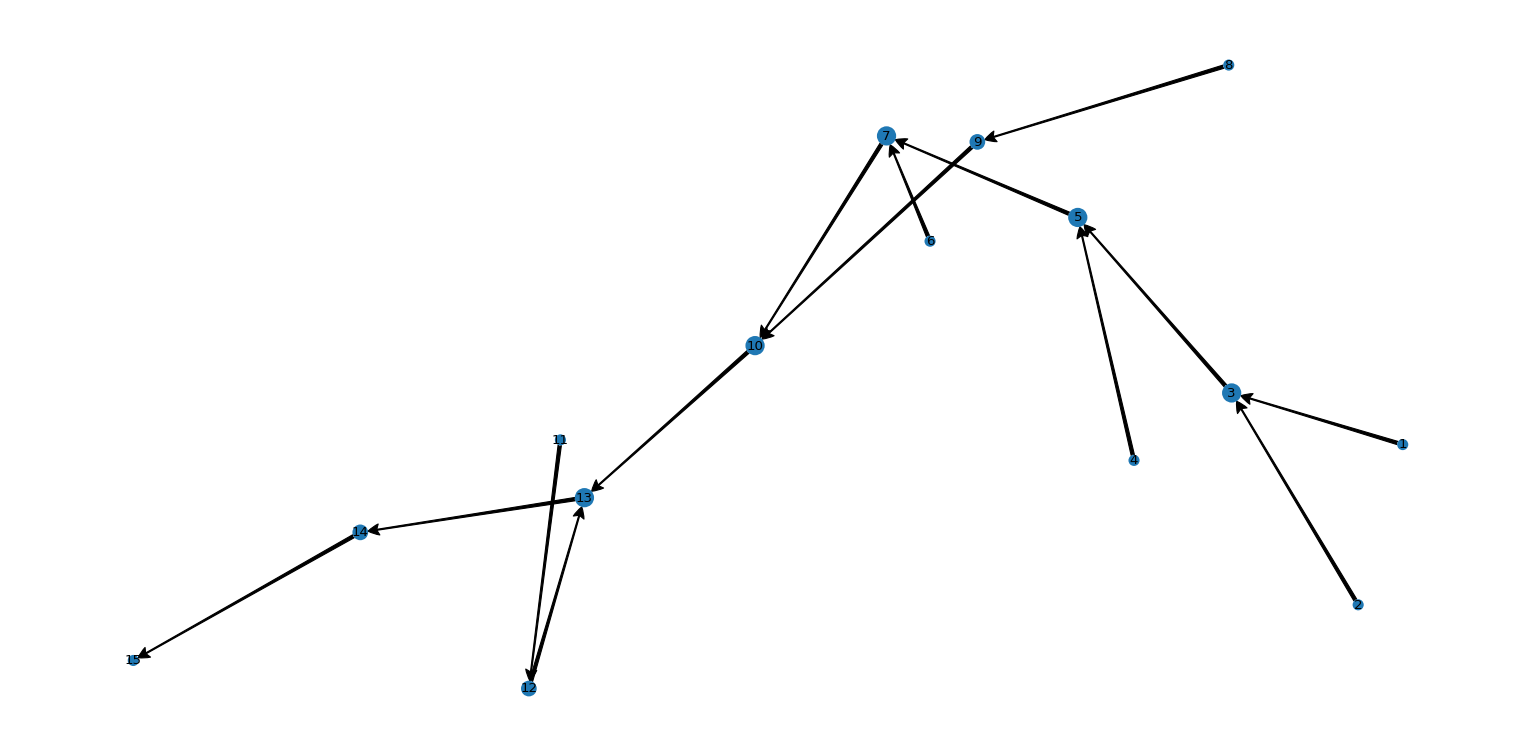}
  \caption{Krishna Basin.}
\end{figure}

\begin{figure}[h]
  \includegraphics[width=\linewidth]{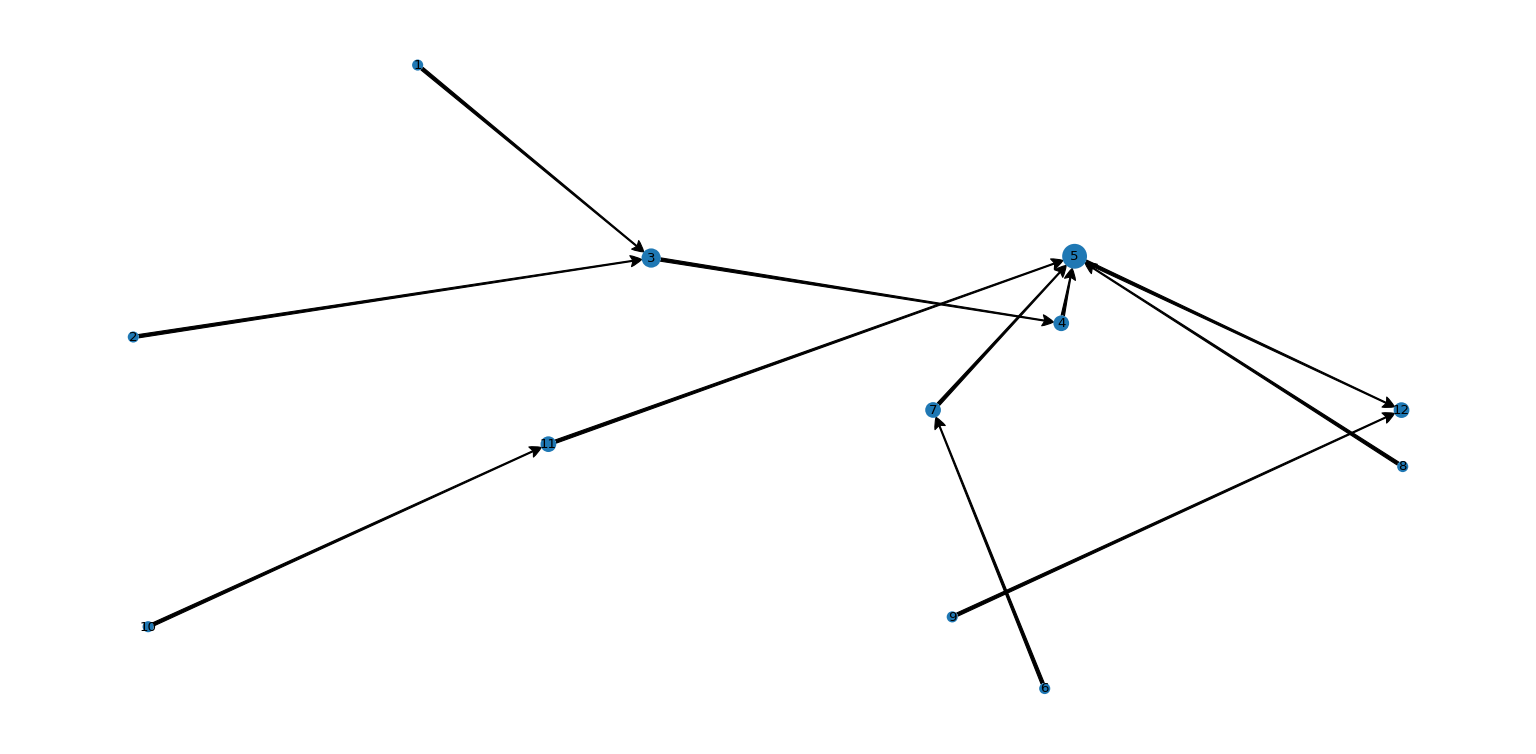}
  \caption{Mahanadi Basin.}
\end{figure}

\begin{figure}[h]
  \includegraphics[width=\linewidth]{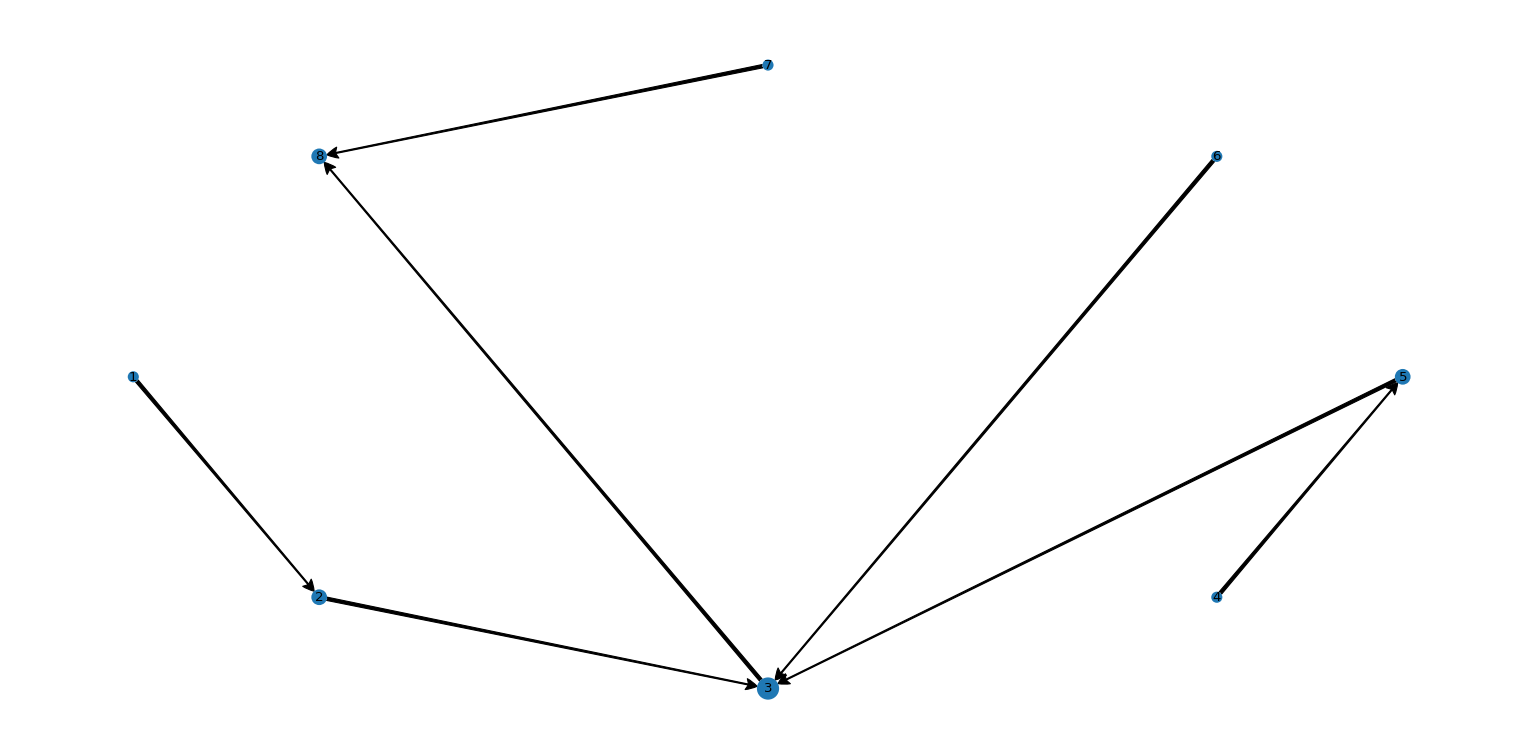}
  \caption{Mahi Basin.}
\end{figure}

\begin{figure}[h]
  \includegraphics[width=\linewidth]{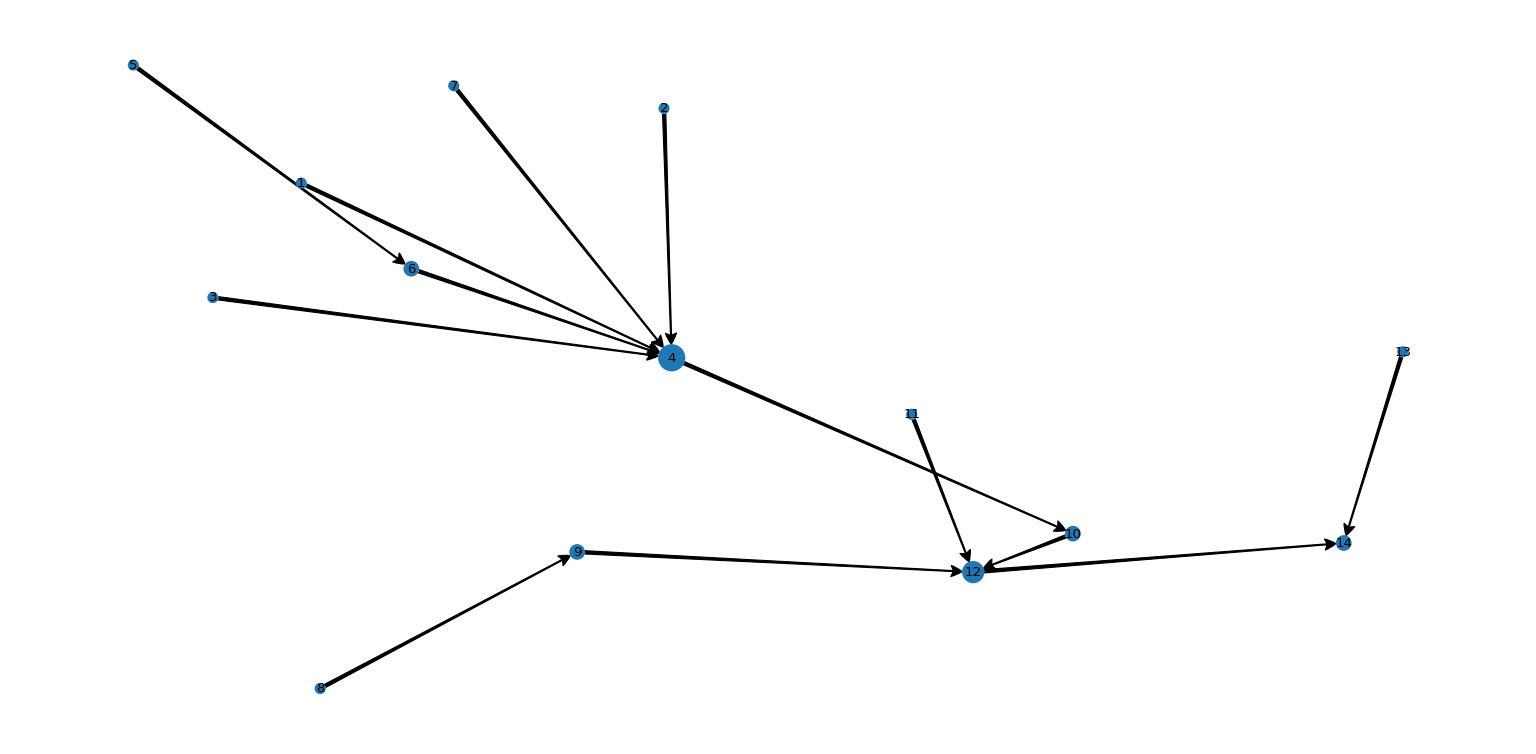}
  \caption{Narmada Basin.}
\end{figure}

\begin{figure}[h]
  \includegraphics[width=\linewidth]{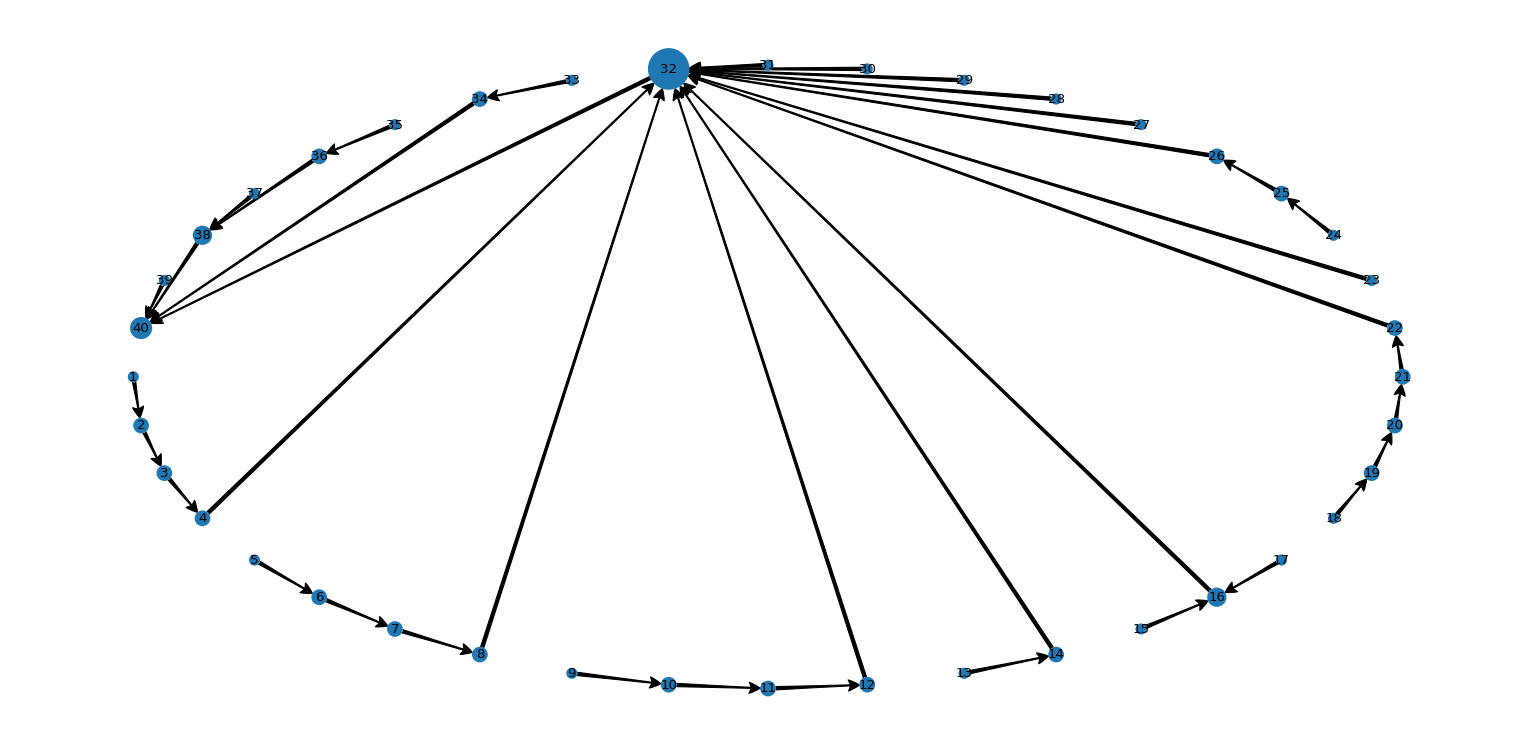}
  \caption{Ganga Basin.}
\end{figure}

\begin{figure}[h]
  \includegraphics[width=\linewidth]{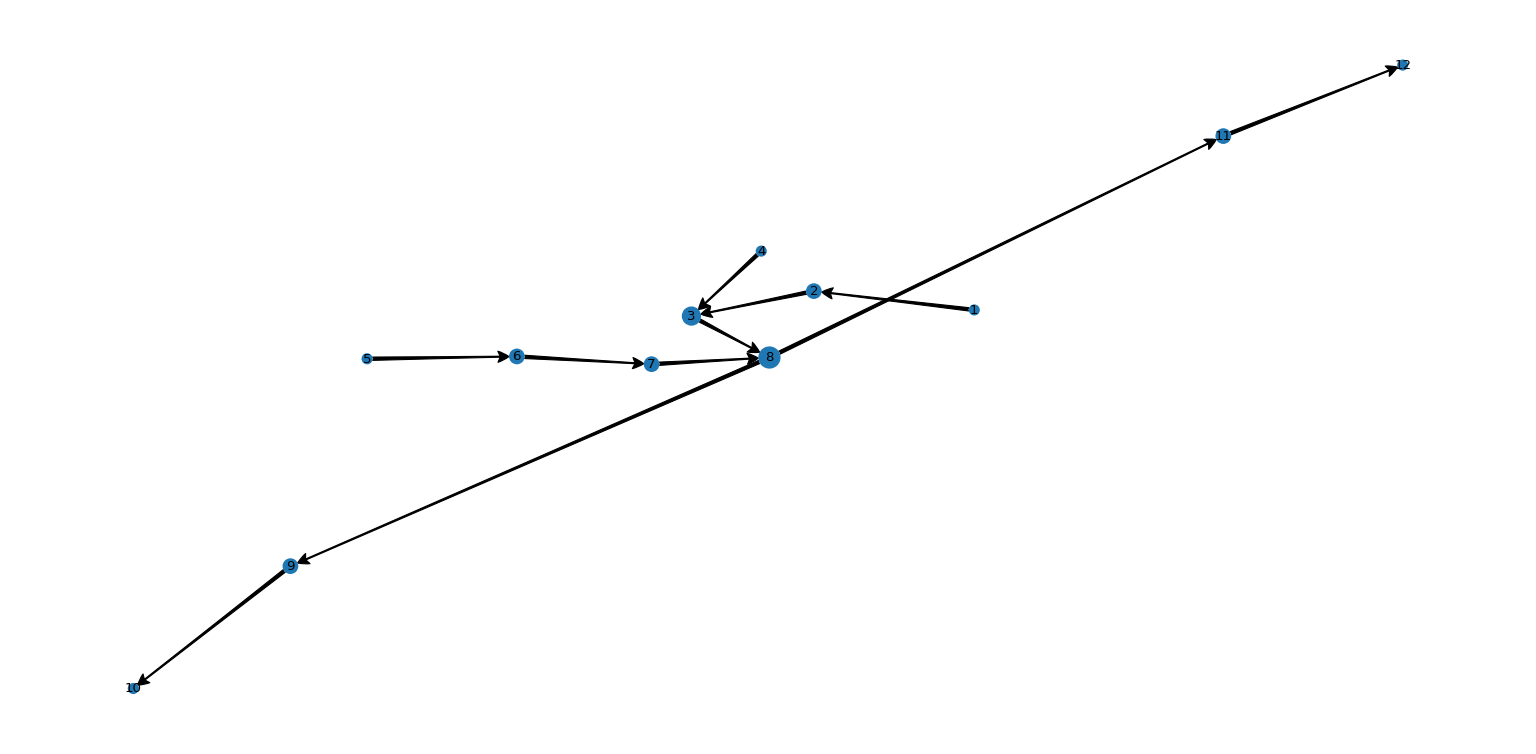}
  \caption{Pennar Basin.}
\end{figure}

\begin{figure}[h]
  \includegraphics[width=\linewidth]{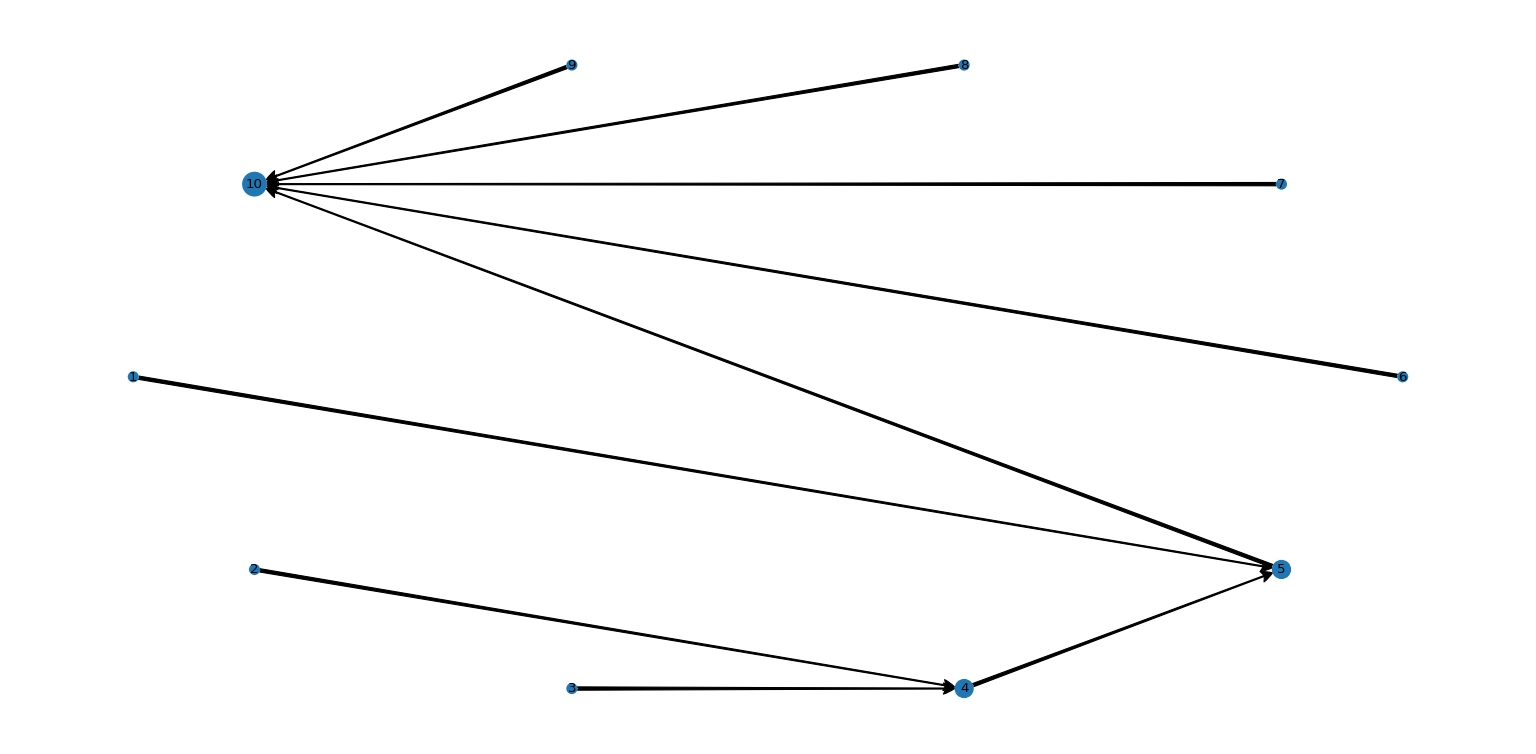}
  \caption{Sabarmati Basin.}
\end{figure}

\begin{figure}[h]
  \includegraphics[width=\linewidth]{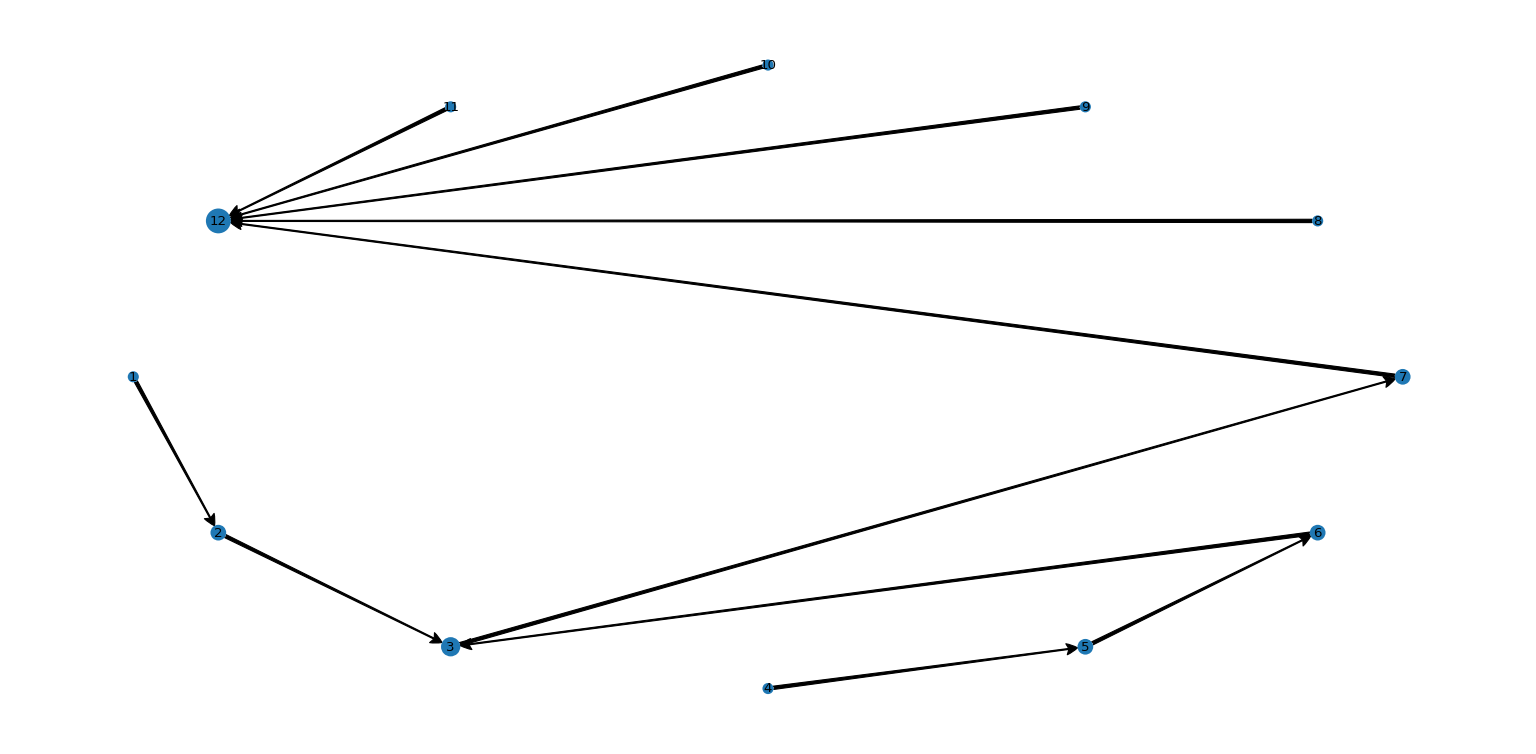}
  \caption{Subarnarekha Basin.}
\end{figure}

\begin{figure}[h]
  \includegraphics[width=\linewidth]{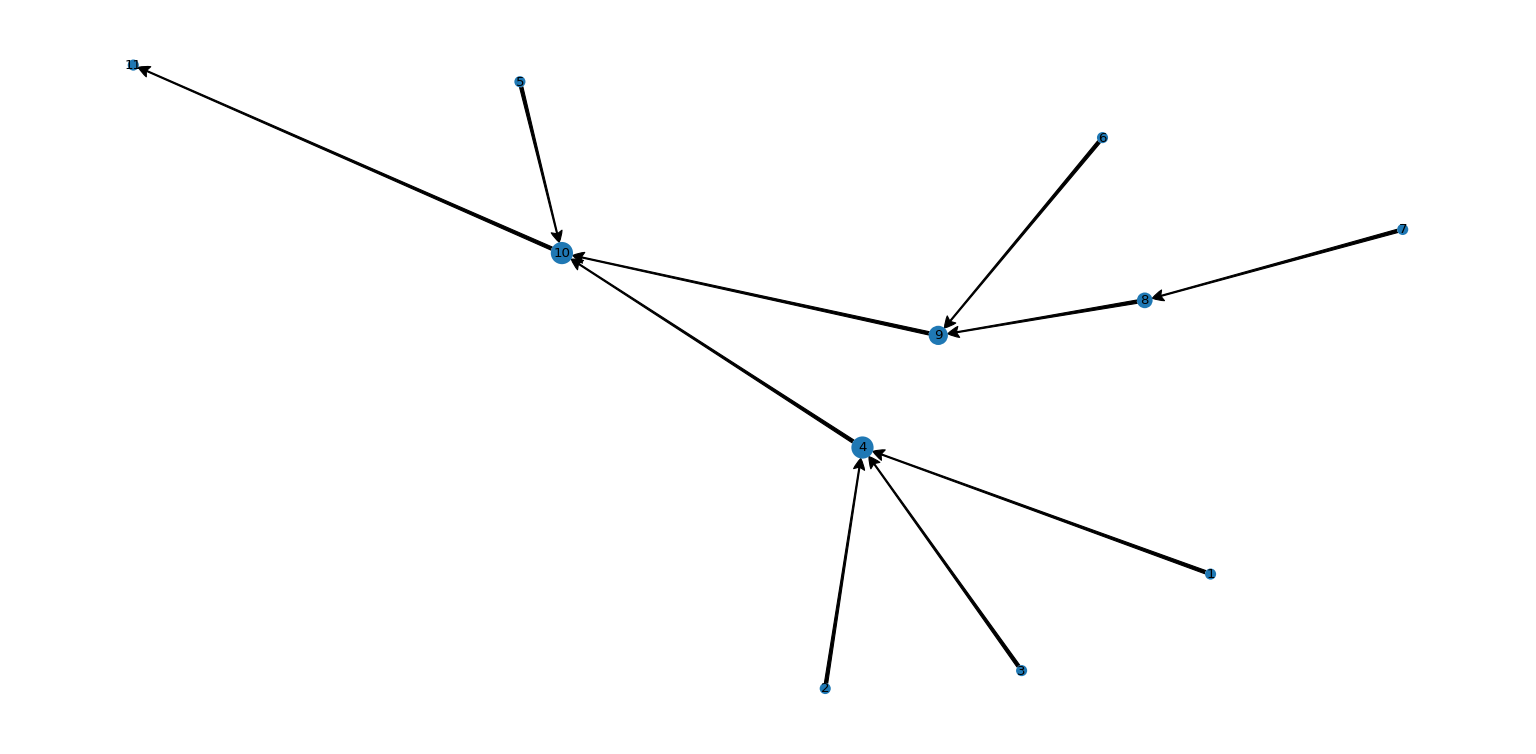}
  \caption{Tapi Basin.}
\end{figure}

\end{document}